\documentclass[conference]{IEEEtran}
\IEEEoverridecommandlockouts
\usepackage{cite}
\usepackage{amsmath,amssymb,amsfonts}
\usepackage{algorithmic}
\usepackage{graphicx,epstopdf}
\usepackage{textcomp}
\usepackage{xcolor}
\def\BibTeX{{\rm B\kern-.05em{\sc i\kern-.025em b}\kern-.08em
    T\kern-.1667em\lower.7ex\hbox{E}\kern-.125emX}}
\begin{document}

\title{Wideband Josephson Parametric Amplifier with Integrated Transmission Line Transformer}

\author{\IEEEauthorblockN{Leonardo Ranzani}
\IEEEauthorblockA{\textit{Quantum engineering and Computing} \\
\textit{Raytheon BBN}\\
Cambridge, Massachusetts, USA \\
leonardo.ranzani@raytheon.com}
\and
\IEEEauthorblockN{Guilhem Ribeill}
\IEEEauthorblockA{\textit{Quantum engineering and Computing} \\
\textit{Raytheon BBN}\\
Cambridge, Massachusetts, USA \\
guilhem.ribeill@raytheon.com}
\and
\IEEEauthorblockN{Brian Hassick}
\IEEEauthorblockA{\textit{Quantum engineering and Computing} \\
\textit{Raytheon BBN}\\
Cambridge, Massachusetts, USA \\
brian.hassick@raytheon.com}
\and
\IEEEauthorblockN{Kin Chung Fong}
\IEEEauthorblockA{\textit{Quantum engineering and Computing} \\
\textit{Raytheon BBN}\\
Cambridge, Massachusetts, USA \\
kc.fong@raytheon.com}
}

\maketitle

\begin{abstract}
We describe a wide-band Josephson Parametric Amplifier (JPA) that is impedance-matched using an integrated compact superconducting transmission line transformer. The impedance transformer consists of two broadside coupled transmission lines configured in a Ruthroff topology which enables a wide matching bandwidth from 2 to 18\,GHz, reducing the input line impedance and the device resonance quality factor by a factor of 4. This enables gain flatness and flexibility in the choice of the amplifier's tuning range. The amplifier has up to 20dB gain, with less than 1\,dB of ripple, 2-3\,GHz gain-bandwidth product and -126\,dBm input 1-dB compression point. Moreover, the device active area fits into a $1\,mm \times 1\,mm$ space, thus easing integration into large quantum systems.\\
\end{abstract}

\begin{IEEEkeywords}
Parametric amplifiers, superconducting microwave devices, Josephson junctions, impedance transformers
\end{IEEEkeywords}

\section{Introduction}

Parametric amplifiers enable the amplification of weak readout signals of quantum devices~\cite{aumentado2020superconducting,lecocq2017nonreciprocal,lecocq2020microwave,bergeal2010phase,chien2020multiparametric,sliwa2015reconfigurable,mutus2013design} to increase readout fidelity and enable quantum process characterization and validation~\cite{devoret2013superconducting}, fast quantum feedback~\cite{riste2013deterministic} and error correction~\cite{campagne2020quantum}. Parametric amplification is typically achieved by modulating a resonant circuit at twice its resonant frequency, which results in the transfer of energy from the modulating drive (pump) to the input signal. Josephson parametric amplifiers (JPAs) use a Superconducting Quantum Interference Device (SQUID) as a tunable inductive element that is modulated by driving the current through an external flux line. JPAs have demonstrated high gain, in excess of 20\,dB and added noise close to the standard quantum limit, or an equivalent input noise of one half of a noise photon. Traveling wave parametric amplifiers (TWPAs) have also been demonstrated, which consist of a long array of Josephson junctions or SQUIDs that is modulated by a pump co-propagating with the input signal~\cite{macklin2015near,zorin2016josephson,bell2015traveling}. TWPAs are characterized by wider bandwidth and higher saturation power than JPAs, at the price of increased circuit complexity and higher noise~\cite{esposito2021perspective}. Alternatively high kinetic inductance materials have also been used to implement traveling wave amplification~\cite{ho2012wideband,ranzani2018kinetic,chaudhuri2017broadband}. An ideal parametric amplifier should have high gain and low noise in order to maximize the quantum efficiency of the measurement chain and at the same time wide bandwidth and high saturation power are desirable for readout multiplexing and high measurement speed. While Josephson Parametric Amplifiers (JPAs) provide near quantum-limited noise performance and high gain, they are based on resonant elements and have a fixed gain-bandwidth product which is set by the resonance linewidth and it is typically a few hundred MHz, resulting in typically less than 20\,MHz of bandwidth at 20\,dB gain. 
Recent work has shown that the bandwidth of a JPA can be further improved beyond the device gain-bandwidth product by using a passive network to impedance match the input RF line to the JPA input impedance~\cite{mutus2014strong,roy2015broadband,naaman2019high}. The matching network needs to: a) reduce the impedance of the input line, thus reducing the resonance Q factor, and b) cancel the reactive component of the JPA input impedance. In this case the shape of the gain profile is in general not Lorentzian and the device gain-bandwidth product ceases to be a constant, being limited only by the Bode-Fano theorem. Previously demonstrated impedance matched parametric amplifier used long tapers~\cite{mutus2014strong,roy2015broadband,grebel2021flux,duan2021broadband} and complex reactive networks, resulting in a large device area. In this work we describe an on-chip wideband Josephson parametric amplifier that uses a transmission line impedance transformer~\cite{ranzani20124,ranzani2013broadband} to widen the amplifier instantaneous bandwidth of the JPA and increase saturation power without sacrificing chip area. Moreover the compact size of the matching network reduces signal reflections and gain ripple, resulting in less than 1\,dB gain variation.

\section{Parametric Amplifier Design}
\begin{figure}
    \centering
    \includegraphics[width=\columnwidth]{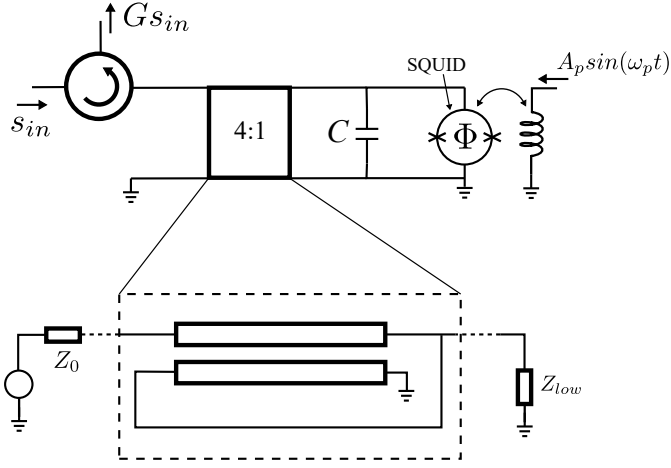}
    \caption{Schematic of the wideband JPA, consisting of a flux-pumped resonant circuit and an integrated impedance transformer, to lower the input line impedance from 50\,Ohm to 12.5\,Ohm. The impedance $Z_{low}$ includes the effect of a short section of microstrip transmision line connecting the transformer to the SQUID. }
    \label{fig:fig1}
\end{figure}
The JPA we designed is shown in Fig.~\ref{fig:fig1}: it consists of a Superconducting Quantum Interference Device (SQUID), shunted by a large capacitor to create a resonance. The SQUID behaves as a tunable inductance that can be modulated by driving the flux through the SQUID loop. We achieve parametric amplificaton by dc-biasing the SQUID and modulating its inductance at the sum of the signal $\omega_s$ and idler $\omega_i$ frequencies $\omega_p=\omega_s+\omega_i\approx 2\omega_s$. As a result, when a small signal at frequency $\omega_s$ is injected at the device input port, an amplified signal is reflected at both $\omega_s$ and $\omega_i$. In the small signal regime, the modulated SQUID inductance is described via the pumpistor model~\cite{sundqvist2013pumpistor,mutus2014strong} as an equivalent frequency-dependent static admittance $Y_A(\omega_s)=1/(j\omega_sL_0)+1/(j\omega_sL_1+X)$, where:

\begin{equation}
\begin{aligned}
L_0 &= L_J/cos(\pi\Phi_{dc}/\Phi_0) \\
L_1 &= -\frac{4L_Jcos(\pi\Phi_{dc}/\Phi_0)}{\pi^2sin^2(\pi\Phi_{dc}/\Phi_0)}\frac{\Phi_0^2}{\Phi_{ac}^2}\\
X &= -\frac{4\omega_s\omega_iL_J^2Y^*_{ext}(\omega_i)}{\pi^2sin^2(\pi\Phi_{dc}/\Phi_0)}\frac{\Phi_0^2}{\Phi_{ac}^2}
\end{aligned}
\label{Eq1:pumpistor}
\end{equation}

In (\ref{Eq1:pumpistor}) $L_J=\Phi_0/2\pi I_c$ is the SQUID inductance at zero bias, $\Phi_0$ is the magnetic flux quantum, $\Phi_{dc}$ the bias flux and $\Phi_{ac}$ the flux modulation amplitude. Note that if $Re[Y_{ext}(\omega_i)] \neq 0$ the real part of $X$ is a negative resistance, which gives rise to gain. Moreover, $Re[X]$ depends on the admittance at the idler frequency seen by the amplifier and therefore impedance matching at both the signal and idler frequencies is required to achieve gain. The gain of the parametric amplifier can be computed by calculating the reflection coefficient:

\begin{equation}
G = \frac{Y_{ext}(\omega_s)-Y_A(\omega_s)}{Y_{ext}(\omega_s)+Y_A(\omega_s)}
\label{Eq2:gain}
\end{equation}

\begin{figure}
    \centering
    \includegraphics[width=0.92\columnwidth]{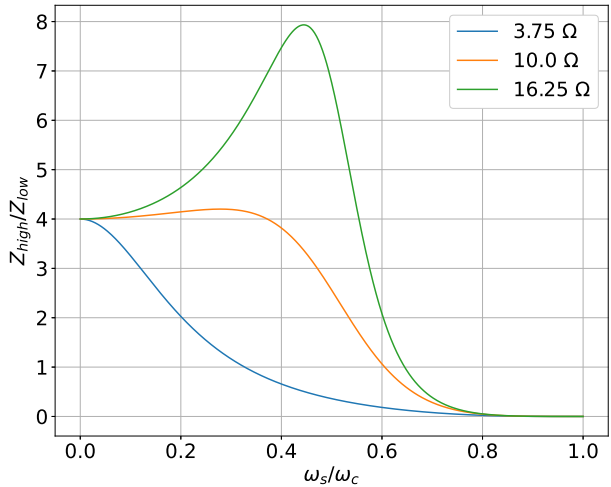}
    \caption{Impedance transformation ratio of the Ruthroff transformer as a function of frequency for different odd-mode impedance values. We obtain a flat transformation ratio for $Z_{oo}=10\,\Omega$  }
    \label{fig:fig2}
\end{figure}

Note that when the environment admittance is real, as is the case when the amplifier is connected to an input transmission line with impedance $Z_0=1/Y_{ext}$, the amplifier gain profile is Lorentzian with a gain-bandwidth product controlled by the Q factor of the bare, unpumped, resonance $Q=\omega_0Z_0C$. Therefore a simple strategy to increase the amplifier bandwidth is to lower the environment impedance $Z_0$, which has the added benefit of increasing bandwidth at a fixed operating frequency without decreasing the Josephson junction critical current and therefore preserves the amplifier saturation power. In our implementation, we use a compact on-chip wideband transformer to reduce the line impedance from 50\,$\Omega$ to 12.5\,$\Omega$ and separately tune the cable length between the circulator and the device to match the reactive part of $Y_A(\omega_s)$. Our transformer, schematically shown in Fig.~\ref{fig:fig1}, consists of a pair of coupled transmission lines connected in parallel at the low impedance end and in series at the high impedance end. This topology, commonly referred as Ruthroff transformer, provides wideband impedance transformation when the length of the transmission lines is much shorter than the signal wavelength~\cite{ruthroff1959some}. We can model signal propagation in the coupled transmission lines as a combination of even and odd modes, corresponding to either equal (even) or opposite (odd) currents flowing into the two lines. We assume that the lines are tightly coupled, so that the differential capacitance between the transmission lines is much greater than the capacitance of each individual line to ground and therefore the even mode characteristic impedance $Z_{0e}$ is much larger than the odd mode impedance $Z_{0o}$. We then approximately compute the impedance at the low-impedance port as~\cite{ranzani20124}:

\begin{equation}
Z_{ext} = \frac{1}{Y_{ext}}= 2Z_{oo}\frac{Z_o\cos\theta-2jZ_{oo}\sin\theta}{4Z_{oo}(\cos\theta+1)-jZ_o\sin\theta},
%Z_{high} = 4Z_{0o}\frac{Z_{low}(\cos\theta+1)+jZ_{0o}\sin\theta}{2Z_{0o}\cos\theta+jZ_{low}\sin\theta}
\label{Eq:Eqimpedance}
\end{equation}

where $Z_o$ is the characteristic impedance of the input line, $\theta=\beta l$ is the electrical length of the coupled lines, with $\beta=\omega_s/v$ and $v$ the odd mode phase velocity. We plot the transformer impedance ratio as a function of frequency in Fig.~\ref{fig:fig2}. The impedance ratio is close to 4 at low frequency, when the electrical length is much shorter than $\pi$ and drops to zero for $\omega_s=\omega_c=2\pi \times v/2l$. Moreover the impedance transformation is approximately flat in frequency when the odd mode impedance is about 20\% of the high impedance $Z_o$. Using Eqs. (\ref{Eq1:pumpistor}-\ref{Eq:Eqimpedance}) we can compute the gain of the parametric amplifier and optimize the transformer design.\\
\section{Transformer design and fabrication}
\begin{figure}
    \centering
    \includegraphics[width=0.9\columnwidth]{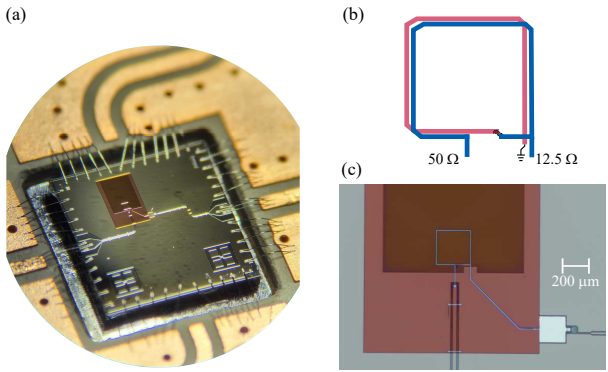}
    \caption{(a) Picture of the fabricated device. The input RF signal is injected from the left port, while the RF pump and dc bias are applied on the right port. (b) Sketch of the impedance transformer consisting of two broadside coupled transmission lines patterned in a ring geometry and connected as in Fig.~\ref{fig:fig1}. (c) Detail of the amplifier active area, including the transformer, parallel plate capacitor and SQUID.}
    \label{fig:fig3}
\end{figure}

\begin{figure}
    \centering
    \includegraphics[width=\columnwidth]{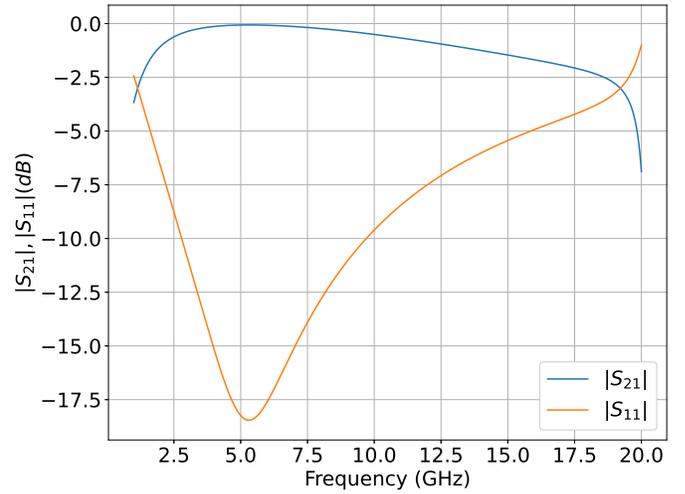}
    \caption{Numerical simulation of the impedance transformer, showing a 3-dB bandwidth of 16~GHz, with better than 0.2~dB insertion loss from 4 to 7.5~GHz.}
    \label{fig:fig4}
\end{figure}

We layout and numerically simulate the transformer in Ansys HFSS. We show a picture of the fabricated device in Fig.~\ref{fig:fig3}. The layout consists of two Nb broadside coupled transmission lines separated by a thin layer (700\,nm) of Silicon nitride deposited via sputtering. The gap between the coupled lines is much smaller than the silicon substrate thickness (500\,$\mu m$), which guarantees a high ratio between even and odd mode impedances $Z_{oe}/Z_{oo} \approx 500/0.7 = 714$ (the ratio between even and odd mode impedances is approximately equal to the ratio between the even and odd mode capacitances). Therefore Eq.~(\ref{Eq:Eqimpedance}) is a good approximation of the transformer impedance ratio. The lines are patterned in a ring geometry to reduce the length of their connection at the transformer low impedance side, which is essential to achieve wide bandwidth. The 50\,$\Omega$ input line is a coplanar waveguide (CPW) patterned on the bottom metal layer. The output line connected to the amplifier is a 12.5\,$\Omega$ microstrip patterned on the top metal layer, which shares the ground plane with the CPW.
We simulated the impedance transformer in Ansys HFSS and plot the results in Fig.~\ref{fig:fig4}. We assumed that the Nb layers are perfect electric conductors, since the kinetic inductance is negligible. We further swept and optimized the device dimensions in the numerical simulation to minimize the insertion loss in the amplifier frequency range between 5 and 7\,GHz.The transformer has better than 3\,dB of insertion loss from 2 to 18~GHz, with less than 0.2\,dB of insertion loss from 4 to 7.5\,GHz, allowing a flexible design of the JPA operating frequency.


\begin{figure}
    \centering
    \includegraphics[width=\columnwidth]{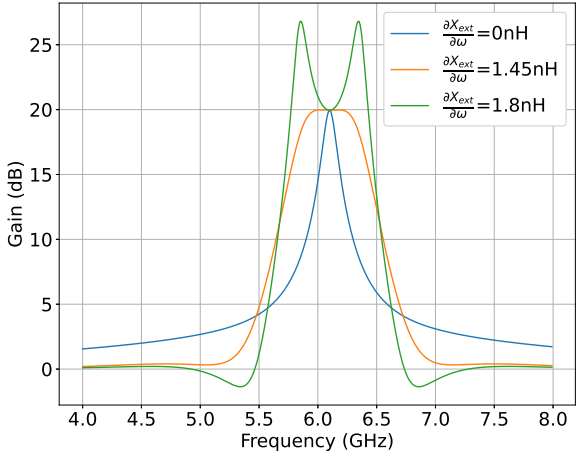}
    \caption{Simulations of the parametric amplifier gain with the matching transformer and different slopes of the reactive part of the external impedance seen by the amplifier. We tune the reactive component by optimizing the length of coaxial cable between the device and the RF circulator.}
    \label{fig:fig5}
\end{figure}

We compute the external admittance $Y_{ext}$ from the full-wave simulations and combine this with Eq.~\ref{Eq1:pumpistor} to simulate the dynamics of the parametric amplifier. While the real part of the transformed impedance $Re(Z_{ext})$ controls the resonance Q factor, the slope of the reactive component $X_{ext} = Im(Z_{ext})$ controls the gain flatness, as shown in Fig.~\ref{fig:fig5}. When $X_{ext}=0$ the gain profile is Lorentzian, it flattens for $X_{ext}>0$ and eventually develops two peaks symmetrically placed around its center frequency. Note that $X_{ext}$ depends on both the transformer parameters as well as the components between the amplifier and the circulator, including in particular the length of cable separating the two~\cite{mutus2014strong}. Reflections from other components are suppressed by the circulator isolation of >40\,dB and therefore have negligible effect on the amplifier gain. At a slope of the reactance $\partial X_{ext}\partial\omega$=1.45\,nH we get a flat gain profile. We simulate that this condition can be reached with a cable length of about 18\,cm between the amplifier and the RF circulator, which tunes out the reactive component of the amplifier impedance seen at the circulator reference plane. This length is of the same order as the length of the mu-metal shield in our dilution refrigerator, however it can be avoided by adding a passive series LC resonator to tune the reactive part of the input impedance~\cite{roy2015broadband}. 

\section{Gain measurements}

\begin{figure}
    \centering
    \includegraphics[width=\columnwidth]{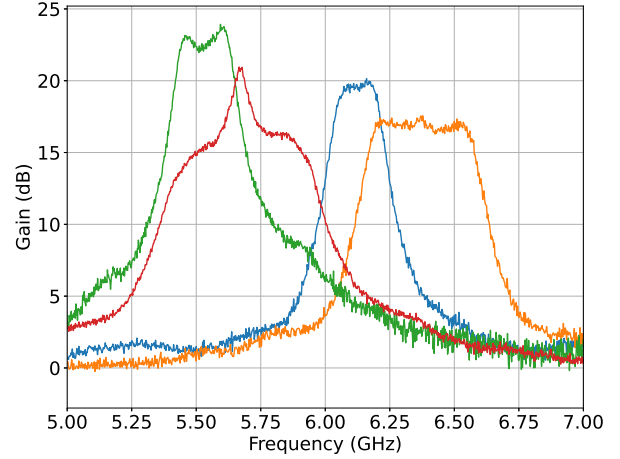}
    \caption{Parametric amplifier gain at two different bias and pump parameter combinations: we measured 20\,dB gain over 200\,MHz of bandwidth and 17\,dB over 450\,MHz.}
    \label{fig:fig6}
\end{figure}

\begin{figure}
    \centering
    \includegraphics[width=\columnwidth]{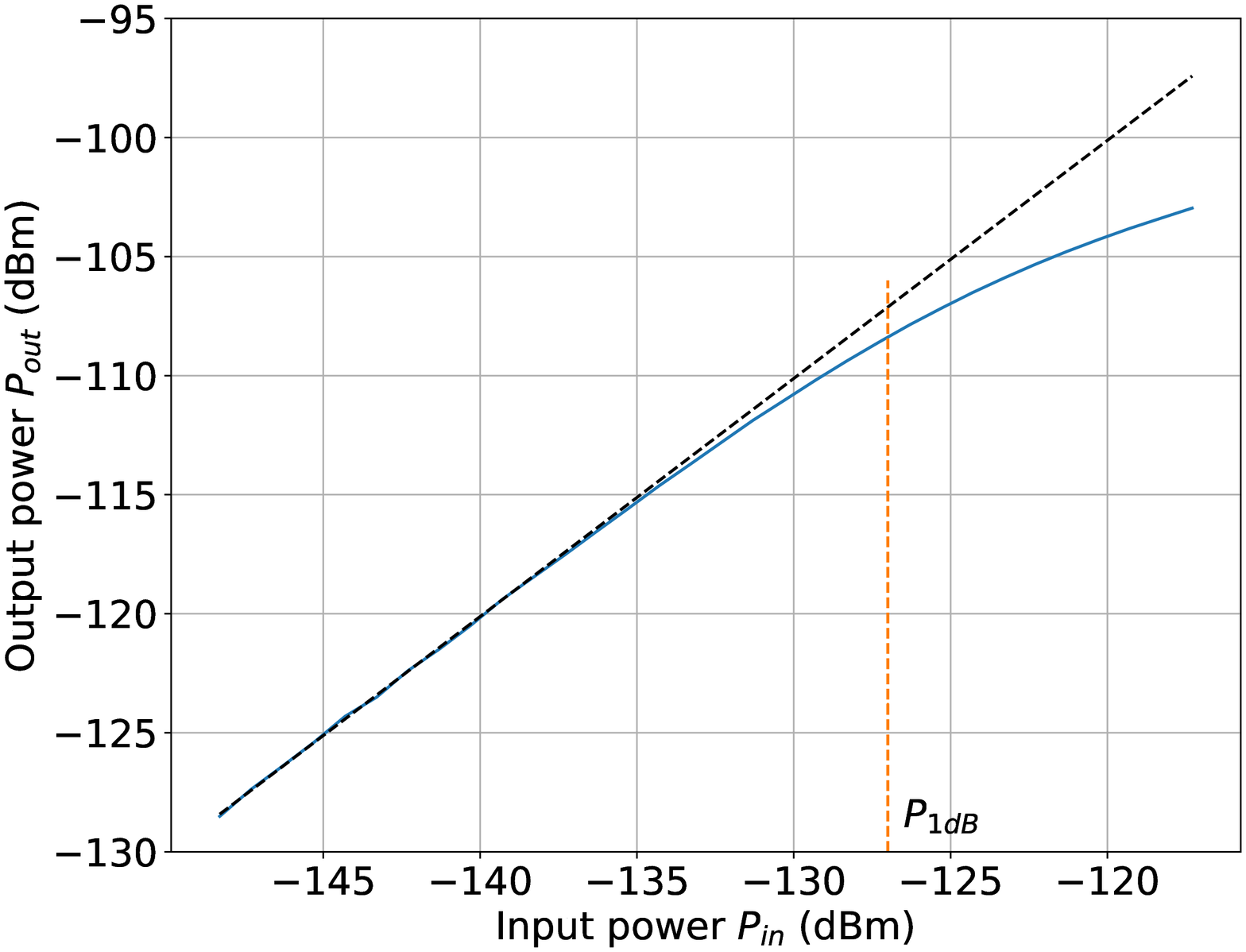}
    \caption{Measured 1-dB compression point of the parametric amplifier at 20\,dB of gain.}
    \label{fig:fig7}
\end{figure}

We wirebond the JPA device to a coplanar waveguide circuit board and packaged it. We then measure it in a dilution refrigerator at a base temperature of 20\,mK. The JPA is a reflection amplifier and therefore we use a commercial cryogenic microwave circulator to inject a weak input signal and measure the amplified output. We attenuate the input signal by 120\,dB and subsequently amplify it by 80\,dB via a cryogenic High Electron Mobility Transistor (HEMT) amplifier operating at 4\,K followed by a RF low noise amplifier at room temperature. We also place a double junction isolator after the circulator to protect the device from 4\,K noise. We use a separate RF line to inject the flux modulation drive (pump) and dc flux bias via a cryogenic bias tee. Finally note that it is important to optimize the cable length between the amplifier and the circulator to achieve optimal impedance match while preventing distortion of the gain profile from standing waves in the cable. 
The measured gain curves are shown in Fig.~\ref{fig:fig6}. We observed wideband gain from 5.2 up to 6.5~GHz by tuning the flux bias and pump frequency. The amplifier is impedance matched over a wide bandwidth of 450\,MHz at 6.3\,GHz with 17\,dB gain, corresponding to a gain-bandwidth product of more than 3\,GHz. The bandwidth decreases to 200\,MHz at 20\,dB gain. The wide bandwidth of the transformer and compact footprint reduces the effect of line reflections on gain flatness, resulting in less than 1\,dB variation in the amplifier gain over the bandwidth, as shown in Fig.~\ref{fig:fig6}. The tuning range of the amplifier covers the frequency range typically used for qubit readout resonators, but smaller than a typical 5-7\,GHz tuning range of a capacitively coupled narrowband JPA~\cite{mutus2014strong}. Moreover the measured bandwidth is also narrower than what our model predicted (about 500\,MHz at 20\,dB, see Fig.~\ref{fig:fig5}). The discrepancy in gain-bandwidth product is consistent with small variations in the external impedance $Z_{ext}$ seen by the amplifier, caused by reflections in the input line, at the circulator and device packaging. Note that at high gain the denominator in Eq.~\ref{Eq2:gain} is close to zero and therefore the gain is very sensitive to small variations in the impedance. We also characterize the amplifier input saturation power by setting the low power gain to 20\,dB and measuring the output signal power as a function of input power. The 1-dB compression point, \text{i.e.} the input signal power where the device gain drops by 1\,dB, is equal to -126\,dBm, as shown in Fig.~\ref{fig:fig7}. Furthermore, the measured input saturation power increases by 1.2\,dBm for every dB of drop in gain. Such a saturation power is comparable to previous implementations of impedance-matched JPAs~\cite{mutus2014strong} for the same level of gain. Further improvement in saturation power can be achieved by replacing the single SQUID with a series of SQUIDs~\cite{naaman2018high}.
\section{Noise measurements}
We perform a hot/cold load measurement of the amplifier noise temperature by connecting the input of the circulator to a variable temperature noise source. The noise source consists of a 50\,$\Omega$ termination anchored to the cold plate at 100\,mK via a low thermal conductivity link and heated by injecting a constant electrical current. The termination is connected to a thermocouple to measure and control its temperature via active feedback. We continuously tune the temperature of the noise source from 400\,mK up to 3\,K and measure the output noise spectral density of the amplified noise $S(\omega,T)$ at frequency $\omega$ and temperature $T$. We then extract the system noise temperature $T_{sys}(\omega)$ by a least squares fit to:

\begin{equation}
    S(\omega,T)=2G\left( \frac{\hbar\omega/k_B}{e^{\hbar\omega/k_BT}-1} +T_{sys}(\omega) \right)
    \label{Eq:noise}
\end{equation}

\begin{figure}
    \centering
    \includegraphics[width=\columnwidth]{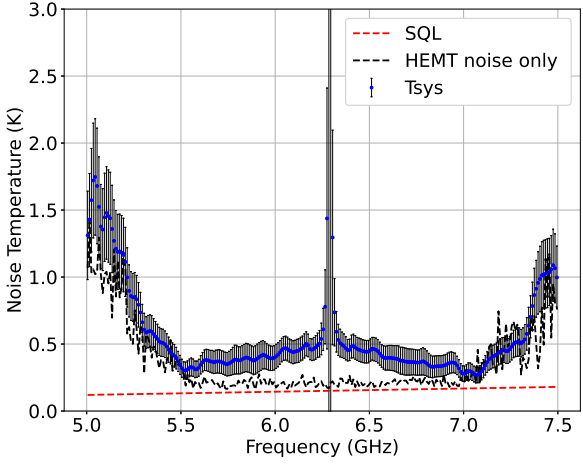}
    \caption{Noise temperature characterization with the parametric amplifier biased at an average gain of 17~dB. The peak at the center frequency is due to measurement errors caused by saturation of the amplifier at high gain, as evidenced by the wider error bars. The standard-quantum-limit (SQL) for a phase-preserving amplifier in the figure corresponds to 1/2 photon $T_{SQL}=(1/2)\hbar\omega/k_B$. We also show the estimated noise contribution of the low noise amplifier at 4K to the total system noise. }
    \label{fig:fig8}
\end{figure}

The measured power spectral density is approximately equal to $S(\omega,T)\approx 2G(T+T_{sys})$ for $T \gg \hbar\omega/k_B$. The factor of 2 in Eq.~\ref{Eq:noise} is due to the fact that we are injecting noise simultaneously at the signal and idler frequencies. We bias and pumped the amplifier to reach a gain of 12-20\,dB and characterize the amplifier noise temperature in Fig.~\ref{fig:fig8}. We measured a noise temperature below 500\,mK  between 5.45\,GHz and 7.25\,GHz, with an average value of 380\,mK. This corresponds to added noise photons $n_{add}=k_BT_{sys}/\hbar\omega$ between 0.8 and 1.5. 

\section{Conclusion}

In this work we have demonstrated a compact wideband Josephson parametric amplifier with 2.5-3\,GHz gain-bandwidth product. The device is integrated on chip and occupies an area of one squared $mm^2$. We use an on-chip Ruthroff transformer to reduce the input line impedance from 50 to 12.5\,$\Omega$ and increase the amplifier bandwidth. The compact design reduces the device area and improves gain flatness compared to traditional tapered line designs. Furthermore we have demonstrated -126\,dBm 1-dB compression point and 1.1 photons of equivalent input noise, making the amplifier suitable for high fidelity qubit readout chains. Further improvements to this design are possible, such as using a chain of SQUIDs to increase the saturation power and use a reactive matching network, in addition to the impedance transformer, to further increase the bandwidth. 

\section*{Acknowledgments}

We acknowledge Thomas Ohki for insightful discussions and Andrew Wagner for fabrication help. This work does not contain any technical data controlled under the ITAR or EAR. This work was performed in part at the Harvard University Center for Nanoscale Systems (CNS); a member of the National Nanotechnology Coordinated Infrastructure Network (NNCI), which is supported by the National Science Foundation under NSF award no. ECCS-2025158.
%The preferred spelling of the word ``acknowledgment'' in America is without 
%an ``e'' after the ``g''. Avoid the stilted expression ``one of us (R. B. 
%G.) thanks $\ldots$''. Instead, try ``R. B. G. thanks$\ldots$''. Put sponsor 
%acknowledgments in the unnumbered footnote on the first page.
\bibliographystyle{IEEEtran}
%\bibliography{jpa_bib}

\end{document}